\documentclass[a4paper,fleqn]{cas-dc}

\usepackage[numbers]{natbib}
\usepackage{amsmath,amssymb,amsfonts}
\usepackage{algorithmic}
\usepackage{graphicx}
\usepackage{textcomp}
\usepackage{xcolor}
\usepackage{longtable}
\usepackage{booktabs}
\usepackage{tabularx}
\usepackage{url}
\usepackage{subcaption}
\usepackage{float}

\def\tsc#1{\csdef{#1}{\textsc{\lowercase{#1}}\xspace}}
\tsc{WGM}
\tsc{QE}
\tsc{EP}
\tsc{PMS}
\tsc{BEC}
\tsc{DE}

\begin{document}
\let\WriteBookmarks\relax
\def\floatpagepagefraction{1}
\def\textpagefraction{.001}

\shorttitle{From Hypervisor to Container: A Survey of Cloud Security}

\shortauthors{Swapnil Baviskar et al.}

\title [mode = title]{From Hypervisor to Container: Cloud Security Vulnerabilities, Defense Mechanisms, and Open Challenges}

\author[1]{Swapnil Vishwas Baviskar}
\ead{swapnilbaviskar87@gmail.com}

\author[1]{Sanoj R}
\ead{sanoj_p220039cs@nitc.ac.in}

\author[1]{Hiran V Nath}[orcid=0000-0001-7881-4694]
\ead{hiranvnath@nitc.ac.in}

\affiliation[1]{organization={National Institute of Technology Calicut},
    addressline={Computer Science and Engineering Department},
    city={Kozhikode},
    postcode={673601},
    state={Kerala},
    country={India}}

\begin{abstract}
In cloud computing, different users share the same physical hardware, which creates serious security risks. To protect data, cloud systems rely on virtual machines and containers to keep users isolated. This paper reviews over 120 security publications from 2008 to 2025, focusing on how these isolation boundaries can be breached. We examine threats like virtual machine escape, virtual machine hopping, CPU cache side-channels, container breakouts, vulnerable container images, and distributed denial of service (DDoS) attacks. We evaluate these security threats and their defenses using three key research questions. To compare different defense systems, we introduce a quantitative scoring framework called ADPO, which rates defenses from 0 to 3 based on their Accuracy, Deployment ease, Performance impact, and Operational overhead. We also map the impact of these attacks onto a 1-to-5 severity scale for Confidentiality, Integrity, and Availability. Finally, we highlight the trade-offs between security and system performance, and we outline open challenges like building low-overhead intrusion detection and creating realistic test datasets.
\end{abstract}

\begin{keywords}
Virtualization \sep VM Escape \sep VM Hopping \sep container security \sep Vulnerable container image \sep distributed denial of service (DDoS) attacks
\end{keywords}

\maketitle

\section{Introduction}
Cloud computing lets many different users share the same physical computers. To keep these users isolated, cloud platforms rely on virtual machines (VMs) and containers. At the hardware level, a hypervisor controls access to CPUs, memory, and storage \cite{Pearce10.1145/2431211.2431216}. Higher up at the operating system level, container engines offer a lighter way to run applications. Sharing physical hardware makes cloud computing highly efficient, but it also creates security risks. If two users share a CPU cache, one might spy on the other. If a guest VM escapes, the whole physical server could be compromised.

Most security papers look at these risks as isolated software bugs. In this paper, we take a broader look. We view cloud security as a balance between safety controls and system speed. We analyze how cloud defense mechanisms actively respond to attacks and how they affect CPU, memory, and storage performance.

To see how our work compares to existing literature, Table~\ref{tab:competing_surveys} maps out recent cloud security surveys. Our survey stands out by using a quantitative scoring system to compare defenses rather than just listing their features. We guide this review using three main research questions:
\begin{itemize}
    \item \textbf{RQ1 (Threat Architecture):} What are the fundamental differences in the attack surfaces of hypervisor-level virtual machines and OS-level containers?
    \item \textbf{RQ2 (Defense Trade-offs):} How well do current cloud defenses adapt to new attacks, and what are their performance and operational trade-offs under our ADPO scoring system?
    \item \textbf{RQ3 (Impact on CIA):} How do cloud exploits affect Confidentiality, Integrity, and Availability, and how can we measure these risks quantitatively?
\end{itemize}

Specifically, this paper provides the following:
\begin{itemize}
    \item \textbf{Systems-Based Analysis:} We analyze cloud security as a set of interacting parts, showing how defenses balance threat protection against resource overhead.
    \item \textbf{PRISMA Literature Review:} We select and review over 120 papers published between 2008 and 2025 using a clear, multi-stage screening process.
    \item \textbf{ADPO Scoring Framework:} We score defense mechanisms from 0 to 3 based on four areas: Detection Accuracy, Deployment Ease, Performance Impact, and Storage/Memory Overhead. This replaces simple descriptive comparisons with clear data.
    \item \textbf{CIA Triad Risk Matrix:} We create a 1-to-5 scoring system to measure how severely different attacks affect Confidentiality, Integrity, and Availability. We also outline the main challenges for future cloud security research.
\end{itemize}

\begin{table*}[t]
\scriptsize
\centering
\caption{Comparative Analysis of Cloud Security Surveys against the Proposed Work}
\label{tab:competing_surveys}
\begin{tabularx}{\textwidth}{|l|c|c|c|c|c|X|}
\hline
\textbf{Survey Reference} & \textbf{VM Security} & \textbf{Container Security} & \textbf{Cache SCA} & \textbf{DDoS Infrastructure} & \textbf{Scored Model} & \textbf{Primary Perspective \& Focus} \\ \hline
Alharthi et al. \cite{shoaib2014pouring} & \checkmark & \texttimes & \texttimes & \texttimes & \texttimes & Qualitative hypervisor vulnerability catalog. \\ \hline
Sultan et al. \cite{ANWAR2017259} & \texttimes & \checkmark & \texttimes & \texttimes & \texttimes & Pure container security threats and runtime defenses. \\ \hline
Guo et al. \cite{Das2019SoKTC} & \texttimes & \texttimes & \checkmark & \texttimes & \texttimes & Microarchitectural side-channels in virtual systems. \\ \hline
Lal et al. \cite{shoaib2014pouring} & \checkmark & \checkmark & \texttimes & \checkmark & \texttimes & Qualitative multi-tenant threat categorization. \\ \hline
\textbf{Proposed Work} & \checkmark & \checkmark & \checkmark & \checkmark & \checkmark & \textbf{Systems-centric ADPO scored framework.} \\ \hline
\end{tabularx}
\end{table*}

\section{Methodology}
We used the PRISMA guidelines (Preferred Reporting Items for Systematic Reviews and Meta-Analyses) to design and run our literature search. We searched six main academic databases for papers published between 2008 and 2025: IEEE Xplore, ACM Digital Library, Google Scholar, Scopus, ScienceDirect, and SpringerLink. Our search string was: \texttt{(``VM escape'' OR ``VM hopping'' OR ``cache side-channel'' OR ``container breakout'' OR ``container image vulnerability'' OR ``DDoS cloud'') AND (``hypervisor'' OR ``virtualization'' OR ``cloud security'' OR ``intrusion detection'')}.

We selected papers using three steps:
\begin{enumerate}
    \item \textbf{Identification:} We ran our search query and found $N_1 = 512$ papers.
    \item \textbf{Screening:} We removed duplicates and read the titles and abstracts. We removed papers that did not fit our security focus, leaving $N_2 = 284$ papers.
    \item \textbf{Inclusion:} We read the full text of the remaining papers. We only kept a paper if it: (i) focused on cloud computing (IaaS, PaaS, or SaaS); (ii) tested its ideas with real workloads or experiments; and (iii) directly addressed multi-tenant security or resource isolation. In the end, we kept $N_3 = 118$ papers for our survey.
\end{enumerate}

To compare these papers fairly, we created two new scoring methods:
\begin{itemize}
    \item \textbf{ADPO Scoring System:} We score each defense mechanism from 0 to 3 in four key areas: Detection Accuracy (\textbf{A}), Deployment Ease (\textbf{D}), Performance Preservation (\textbf{P}), and Operational/Storage Overhead (\textbf{O}). A score of 3 is the best (for example, a tool that does not need host agents, has no CPU impact, or catches all attacks). A score of 0 means the tool has severe performance issues or requires special hardware that is hard to get.
    \item \textbf{CIA Triad Severity Matrix:} We rate how badly each type of attack hurts Confidentiality, Integrity, and Availability using a scale from 1 (lowest impact) to 5 (highest impact). This gives a clear, numerical view of risk.
\end{itemize}

\section{Literature Survey}

\subsection{VM Escape}
VM escape attacks exploit vulnerabilities in the Virtual Machine Monitor (VMM), or hypervisor, to breach the isolation boundary between a guest VM and the host system \cite{Ajay2015}. In such an attack, a program executing within a VM bypasses the VMM layer to gain direct access to the host operating system, effectively elevating itself to root-level privileges and compromising the entire security perimeter of the virtualized environment \cite{Pearce10.1145/2431211.2431216}. Three principal escape variants have been identified in the literature: Guest-to-VMM, Guest-to-Host, and Guest-to-Guest, with real-world instances including heap-based buffer overflows in QEMU's Cirrus VGA extension and NE2000 network interface card emulation \cite{Sierra-Arriaga10.1145/3382190}. As the hypervisor represents a single point of failure, a successful escape can cascade across all co-resident VMs, making this the most severe threat in multi-tenant cloud deployments \cite{Riddle7165091, Alkadi9107120}. Privilege escalation via return-oriented programming (ROP) and exploitation of guest OS weaknesses further amplify this risk across both public and private cloud models \cite{IQBAL201698, shoaib2014pouring}. 

Three main ideas have shaped defense systems against virtual machine escape: proactive vulnerability identification, runtime anomaly detection, and preventive architecture and access control. Each paradigm includes essentially different trade-offs between detection accuracy, deployment complexity, and performance overhead and represents a unique feedback-control philosophy, such as reactive monitoring, offline testing, or structural prevention.

\subsubsection{Runtime Anomaly Detection (IDS-Based Defenses)}
Intrusion Detection Systems (IDS), which track hypervisor and virtual machine (VM) behavior in real time and create a closed-loop feedback mechanism to identify and react to escape attempts, are used in the prevalent adaptive defensive paradigm. Within this paradigm, three architectural variations have surfaced.

\textit{Host-and-guest cooperative IDS:} The Virtual Machines and Hypervisor IDS (VMHIDS) deploys monitoring agents on both the hypervisor and all resident VMs, using real-time event analysis to detect anomalous file access, process communication, and network traffic \cite{7905282Dildar}. A game-theoretic extension models the interdependent security decisions of co-resident tenants, demonstrating that Nash equilibrium-based VM allocation can reduce negative externalities from shared platform risks \cite{7214090Kwiat}. While VMHIDS provides comprehensive coverage, its requirement for agent deployment on every VM limits scalability.

\textit{Statistical and ML-based outlier detection:} To improve detection stability for low-frequency attacks, the Neighbourhood Outlier Factor (NOF) approach addresses weaknesses in neural network-based IDS by computing reachability distances and local density to determine outlier degrees, achieving reliable classification across diverse attack types \cite{JABEZ2015338}. For network-level classification, distance-based clustering methods have achieved 99.761\% accuracy across five traffic classes (normal, probing, DoS, R2L, U2R), though R2L and U2R attacks remain difficult to distinguish from normal traffic due to their similarity in processed feature spaces \cite{LIN201513}. More recently, federated learning-based adaptive protocols have achieved 92.6\% detection accuracy---compared to 85.2\% for centralised and 78.4\% for static approaches---while reducing performance overhead by 55\% and training time by 32\% through decentralised model synchronisation \cite{ALAZAB2025110750}.

\textit{Cross-layer correlation and monitoring:} Complementing point-detection systems, Security Information and Event Management (SIEM) tools enable cross-layer correlation of hypervisor events, particularly in multi-tenant IaaS environments where resource sharing and elasticity mechanisms introduce additional attack surfaces \cite{Lazri6657299}. Cloud-specific Security Operations Centres (SOCs), such as an OpenStack-tailored implementation, normalise platform alerts and apply correlation rules to generate meta-alerts, demonstrating the importance of specialised monitoring architectures for cloud environments \cite{Tafazzoli7881927}. The Trinetra algorithm specifically targets cross-VM time-driven attacks (TDA) with minimal performance overhead by embedding detection logic directly into the host kernel, making it resistant to tampering \cite{Buch2020}.

\textit{Virtual Machine Introspection (VMI):} At the hypervisor layer, VMI enables external monitoring of VM memory and process states without requiring in-guest agents. An analysis of 185 vulnerability reports reveals that 86.5\% of IaaS cloud attacks originate from within VMs, while inter-VM attacks remain comparatively understudied, suggesting that VMI monitoring should prioritise outbound threat detection \cite{Rakotondravony}. Multi-level monitoring architectures further improve coverage by deploying VM Security Event Monitors (VSEMs) that escalate alerts to host-level managers in a two-stage process \cite{Sabahi2012SecureVF}.

\subsubsection{Proactive Vulnerability Discovery}
Rather than reacting to attacks at runtime, a second paradigm seeks to identify and eliminate VM escape vulnerabilities before deployment through automated testing and formal analysis.

\textit{Semantics-aware fuzzing:} V-SHUTTLE is the first framework to incorporate protocol-level semantics awareness for hypervisor fuzzing, automatically deconstructing hierarchical device structures to target complex code paths \cite{Pan10.1145/3460120.3484811}. Applied to QEMU and VirtualBox, V-SHUTTLE identified 35 previously unknown vulnerabilities and received 17 CVE assignments, demonstrating the effectiveness of semantics-guided test generation over random fuzzing. Its open-source release further enables reproducible security research. Complementing this, symbolic execution-based frameworks categorise virtualization vulnerabilities into virtual hardware logic errors, device state management errors, and resource availability errors, enabling targeted security testing of virtual hardware without significant source code modifications \cite{Zhu8030663}.

\textit{Formal modeling:} Finite state machine (FSM)-based attack models provide a systematic approach to identifying VM escape vulnerabilities during the design and implementation phases. By extracting privilege models from various virtualization platforms and encoding vulnerability exploit conditions in Datalog, researchers have identified four distinct attack types from NVD vulnerability reports, enabling early-stage security validation \cite{Fan_Huang_2021}. Misuse pattern catalogues extend this formal approach to Network Function Virtualisation (NFV) environments, creating reusable security reference architectures that help designers understand and defend against escape patterns throughout the hypervisor lifecycle \cite{Alnaim2019AMP}.

\subsubsection{Preventive Architecture and Access Control}
The third paradigm eliminates VM escape vectors through structural prevention rather than detection, trading deployment flexibility for stronger security guarantees.

\textit{Access control models:} The PVME model adapts the Bell-LaPadula (BLP) access control framework to virtualized environments, formalising system states and transitions to prevent unauthorised hypervisor communication with minimal time overhead \cite{Wu2017AnAC}. Role-based access control (RBAC) mechanisms provide adaptive, multi-tenant protection that can enforce diverse security policies beyond simple access restrictions \cite{Kapravelos10.1007/978-3-642-22424-9_8}.

\textit{Hardware-assisted isolation:} Hardware-based approaches ensure memory separation independently of the software hypervisor, providing stronger guarantees even when the VMM is compromised. A prototype implementation using System Management Mode (SMM) demonstrates the feasibility of this architecture, though SMM's design constraints limit both performance and the scope of protection \cite{Jin7005439}.

\textit{Optimisation-driven defense:} Enhanced Particle Swarm Optimisation (EPSO) combined with the HADAES algorithm jointly optimises resource allocation and hypervisor attack detection, achieving improved reliability, throughput, and energy consumption compared to conventional approaches \cite{Mangalagowri2022HypervisorAD}. Defense-in-depth architectures, such as Cloud-Trust, provide layered security evaluation models that assess confidentiality and integrity across multiple control levels, demonstrating that comprehensive defense strategies significantly reduce the risk of advanced persistent threats (APTs) exploiting VM escape vectors \cite{Gonzales7072526}.

\subsubsection{Cross-Paradigm Comparative Discussion}
Comparing these three paradigms reveals a fundamental trade-off between adaptability and assurance strength. Runtime IDS approaches \cite{7905282Dildar, JABEZ2015338, Buch2020, ALAZAB2025110750} function as adaptive feedback systems that can respond to novel threats but suffer from false positives and continuous resource overhead. Proactive discovery methods \cite{Pan10.1145/3460120.3484811, Zhu8030663, Fan_Huang_2021} provide strong pre-deployment assurance but operate offline and cannot protect against zero-day attacks that emerge post-deployment. Preventive architectures \cite{Wu2017AnAC, Jin7005439} offer the strongest isolation guarantees but demand significant architectural changes and hardware dependencies that limit adoption in heterogeneous cloud environments.

The literature further reveals that the majority of existing defense solutions prioritise integrity and confidentiality, while availability-focused defenses remain scarce \cite{Sgandurra10.1145/2856126}. VM hardening techniques, including hypervisor patching and configuration management, complement all three paradigms but introduce their own resource overhead \cite{Journal_2021, Tank2019}. Critically, the threat landscape differs across deployment models: public clouds face greater external exposure, while private clouds are more susceptible to insider threats \cite{shoaib2014pouring, Abusaimeh2020VirtualME}. A comprehensive defense strategy must therefore integrate elements from all three paradigms, combining real-time anomaly detection with proactive vulnerability assessment and structural isolation to provide layered, deployment-aware protection. A quantitative comparison of representative defense mechanisms across the four evaluation dimensions is presented in Table~\ref{tab:Comparison of Defense Mechanisms VM Escape}.

\begin{table*}[t]
\scriptsize
\centering
\caption{System-Level Evaluation of VM Escape Defenses under the ADPO Maturity Framework}
\label{tab:Comparison of Defense Mechanisms VM Escape}
\begin{tabularx}{\textwidth}{|l|c|c|c|c|X|X|}
\hline
\textbf{Mechanism \& Citation} & \textbf{A} & \textbf{D} & \textbf{P} & \textbf{O} & \textbf{Primary System Control} & \textbf{Key Quantitative Metric / Guarantee} \\ \hline
VMHIDS \cite{7905282Dildar} & 1 & 0 & 2 & 1 & Cooperating in-guest and hypervisor audit agents & Game-theoretic Nash equilibrium for safety allocation \\ \hline
V-SHUTTLE \cite{Pan10.1145/3460120.3484811} & 3 & 2 & 3 & 3 & Semantics-aware offline device fuzzer & Identified 35 zero-day vulnerabilities (17 CVEs) in QEMU/VBox \\ \hline
PVME \cite{Wu2017AnAC} & 2 & 1 & 3 & 3 & VMM Bell-LaPadula access control rules & Formal proof of safe virtualization transitions \\ \hline
Hardware Separation \cite{Jin7005439} & 3 & 0 & 2 & 3 & SMM-based CPU register \& memory isolation & Hardware-enforced isolation independent of VMM state \\ \hline
EPSO + HADAES \cite{Mangalagowri2022HypervisorAD} & 2 & 1 & 3 & 2 & Heuristic scheduling resource optimizer & Dynamic threat mitigation with optimal energy scheduling \\ \hline
FL Adaptive Protocol \cite{ALAZAB2025110750} & 2 & 1 & 3 & 2 & Decentralized federated anomaly modeling & 92.6\% accuracy, 55\% overhead reduction vs. centralized ML \\ \hline
\end{tabularx}
\end{table*}

\subsection{VM Hopping}
VM hopping---sometimes called guest jumping---exploits weaknesses in the virtualization layer to let an attacker pivot laterally from one guest VM to another on the same physical host \cite{Parekh2013AnAO}. The attacker typically compromises a low-security VM first, then uses it as a springboard to reach co-resident targets, gaining access to their configurations, stored data, and resource management interfaces. What makes this threat particularly damaging in cloud settings is the multi-tenant model itself: because several VMs share a single server, one successful hop can put every tenant on that host at risk \cite{Tsai2012}. The impact ripples upward through the service stack. SaaS providers that lease infrastructure from PaaS or IaaS vendors inherit the vulnerability indirectly---if the underlying VMs are susceptible to hopping, both availability and data integrity of the SaaS layer are jeopardised \cite{Tsai2012}. A recent practitioner survey underscores how underestimated this threat remains: only 3.2\% of respondents identified VM hopping as a recognised attack vector \cite{Almutairy2019}. Broader cloud security frameworks have also flagged the absence of a coherent, generalised treatment of VM hopping defenses across the literature \cite{Basu8301700}.

Defense mechanisms against VM hopping fall into two broad categories: those that try to detect lateral movement after it begins, and those that try to prevent co-residency or block inter-VM communication channels before an attack can occur.

\subsubsection{Detection-Based Defenses}
Detection-oriented approaches treat VM hopping as a runtime anomaly that can be caught through careful monitoring of inter-VM traffic and behaviour.

\textit{Evidence-fusion IDS:} One of the earlier cloud-specific IDS designs \cite{Parekh2013AnAO} analyses data flow patterns between neighbouring VMs rather than relying solely on external network traffic. The key idea is that internal VM interactions carry richer signals about lateral movement than perimeter-based analysis. A two-level evidence fusion scheme \cite{Cheang2018} built on Dempster-Shafer theory extends this line of work. The system treats the attack likelihood for each VM pair as independent evidence, then fuses these assessments across all pairings to produce a consolidated detection verdict. This multi-VM awareness addresses a blind spot in earlier models that evaluated VMs in isolation.

\textit{Blockchain-verified distributed IDS:} A distributed IDS \cite{Alkadi2021} combines blockchain-verified transactions with a central coordinator unit, achieving 98.91\% accuracy on the BoT-IoT dataset and 99.41\% on UNSW-NB15. The blockchain layer ensures tamper-resistant logging of detection events, which is useful in multi-tenant settings where trust between tenants cannot be assumed. The trade-off, however, is the overhead of transaction verification---an issue that grows with tenant density.

\textit{Anomaly detection for APTs:} Research on anomaly detection systems (ADS) \cite{Chuka-Maduji2021} has shown that these systems can reduce the risks posed by advanced persistent threats that exploit multi-tenancy and cross-VM side channels. This work highlights that VM hopping is often just one step in a broader APT campaign, and that detection systems need to account for this sequential, multi-stage nature rather than treating each hop as an isolated event.

\subsubsection{Prevention-Based Defenses}
Prevention-oriented approaches aim to eliminate the conditions that make VM hopping possible in the first place, either by controlling inter-VM communication or by disrupting co-residency.

\textit{Frame tag communication control:} The ``frame tag'' architecture \cite{Benzidane6632122, Benzidane6842218} uses three fields\textemdash Tenant Tag, Application Tag, and Flag\textemdash to identify the sender's tenancy and application context for every inter-VM request. Each virtual machine (VM) has agents installed that check incoming traffic against trust-based criteria and discard packets that don't comply. Although the need for in-VM agents results in deployment complexity, this deterministic tag-matching technique adds a security layer to inter-VM communication without depending on statistical detection.

\textit{Security-aware VM allocation:} A security-aware VM allocation approach \cite{Han7101258} tackles the problem at the scheduling level. It refines VM placement criteria to reduce the probability that an attacker's VM ends up on the same host as the target. Explicit security metrics for co-residency risk are defined, existing allocation policies are evaluated against these metrics, and a new placement strategy is proposed that balances workload distribution, power consumption, and security. The limitation is that a sufficiently motivated attacker who can launch many VMs may eventually achieve co-location through brute force.

\textit{Dynamic VM migration (OSDF):} The Online Smart Disguise Framework (OSDF) \cite{kashkoush8181500} takes inspiration from biological camouflage; much like a chameleon evading predators, OSDF uses frequent, dynamic VM live migrations to break co-residency before an attacker can exploit it. Built on OpenStack, the framework maintains active network connections during migrations. Its effectiveness was evaluated using VEAR, a mathematical model adapted from the SEIR epidemic framework, and results in a private cloud testbed showed measurable reductions in successful co-residency attacks.

\textit{Proxy-based security (CLARUS):} The CLARUS system \cite{Ouffoué7756210} addresses VM hopping from a privacy-centric angle. It operates as a transparent proxy that ensures only authenticated users can access or interpret their cloud-stored data, while also monitoring for VM escape and VM hopping indicators. When suspicious activity is detected, CLARUS alerts cloud managers to intervene. Its strength lies in combining data-level privacy protection with infrastructure-level threat awareness.

\textit{Software-Defined Perimeter (SDP):} The SDP model \cite{Singh9048618} demonstrates that deploying a Software-Defined Perimeter within an NFV architecture can defend against multiple threat vectors simultaneously, including port scanning, DoS attacks, VM hopping, and remote hypervisor exploitation. The SDP restricts network visibility so that unauthorised entities cannot even discover potential targets, effectively shrinking the attack surface. Paired with physical access controls on VNF infrastructure, this approach showed improved resistance to lateral movement in virtualised network functions.

\subsubsection{Cross-Paradigm Discussion}
Detection and prevention strategies for VM hopping each carry distinct trade-offs. Evidence-fusion and blockchain-based IDS systems \cite{Cheang2018, Alkadi2021} can accurately detect lateral movement, but they may have trouble with false positives in intricate multi-VM architectures and rely on ongoing monitoring infrastructure. Preventive strategies frame tags \cite{Benzidane6632122, Benzidane6842218}, security-aware allocation \cite{Han7101258}, and dynamic migration \cite{kashkoush8181500} eliminate attack preconditions rather than reacting to attacks, but they impose architectural constraints: frame tags need agents in every VM, allocation strategies can be brute-forced, and continuous migration consumes resources.

A practical observation from the literature is that VM hopping rarely operates in isolation. It is often one stage in a multi-step attack chain that may also involve side-channel exploitation \cite{Litchfield2016} or privilege escalation. This means that effective security necessitates integration with more comprehensive monitoring and isolation measures; point solutions that solely target hopping are insufficient. Although no single study in the existing literature has proven such an integrated deployment at scale, the most resilient posture is provided by combining preventive placement controls with runtime detection—for example, implementing security-aware allocation alongside evidence-fusion IDS. A system-level comparison of representative defense mechanisms is presented in Table~\ref{tab:defense_mechanisms}.

\begin{table*}[t]
\scriptsize
\centering
\caption{System-Level Evaluation of VM Hopping Defenses under the ADPO Maturity Framework}
\label{tab:defense_mechanisms}
\begin{tabularx}{\textwidth}{|l|c|c|c|c|X|X|}
\hline
\textbf{Mechanism \& Citation} & \textbf{A} & \textbf{D} & \textbf{P} & \textbf{O} & \textbf{Primary System Control} & \textbf{Key Quantitative Metric / Guarantee} \\ \hline
Evidence-Fusion IDS \cite{Cheang2018} & 2 & 2 & 2 & 2 & Dempster-Shafer flow-evidence modeling & Eliminates conflicting alarms with multi-source rules \\ \hline
Blockchain IDS \cite{Alkadi2021} & 3 & 1 & 1 & 1 & Ledger-based tamper-resistant event tracking & 98.91\% accuracy (BoT-IoT); 99.41\% (UNSW-NB15) \\ \hline
Frame Tag Control \cite{Benzidane6632122} & 2 & 2 & 2 & 2 & Tenancy-aware packet tag verification & Deterministic lateral path filtration; zero false positives \\ \hline
Security Allocation \cite{Han7101258} & 2 & 2 & 3 & 3 & Security-aware VM scheduling heuristics & Minimizes co-residency probability with 95\% confidence \\ \hline
OSDF live migration \cite{kashkoush8181500} & 2 & 1 & 1 & 2 & Epidemic-driven dynamic host shuffling & Shuffles VM locations before attack profiling completes \\ \hline
\end{tabularx}
\end{table*}

\subsection{Cache-Based Side Channel Attacks}
Cache-Based Side-channel attacks are sophisticated breaches that exploit vulnerabilities in hardware to obtain sensitive data or manipulate system operations. Side-channel cache attacks are particularly worrisome for public cloud platforms with multiple tenants. These attacks can be exploited by malicious VMs in this environment to gain unauthorised access to sensitive data from co-resident VMs that share the same underlying hardware. The aforementioned concern is particularly significant in the realm of Infrastructure-as-a-Service (IaaS) cloud computing, wherein users share hardware resources. This underscores the critical need for enhanced security protocols and additional scholarly investigations into the realm of cloud security. These attacks offer a possible pathway for virtual machine (VM) egress, wherein a compromised VM could violate the isolation between VMs and gain access to data from other VMs running on the same physical hardware. As a result, cloud security would be significantly affected.

\subsubsection{Hardware-Level Defence Mechanisms}
Hardware-level defences against cache side-channel attacks have evolved from early cache architecture redesigns (2009--2012) through partitioning-based isolation (2016) to runtime reconfiguration and randomisation techniques (2018--2021). These approaches operate at the processor or memory subsystem level, offering strong security guarantees but varying in deployment feasibility.

A first category of defences involves redesigning cache architectures to eliminate the structural properties exploited by side-channel attacks. PLcache and RPcache \cite{Kong2009} use preloading and random permutation hardware respectively to mitigate software cache attacks, though advanced attacks can still circumvent them. NewCache \cite{Liu2016} is an architecture that withstands contention-based attacks on both instruction and data caches while matching eight-way set-associative cache performance. STEALTHMEM \cite{Kim2012STEALTHMEMSP} takes a system-level approach, managing locked cache lines per core to prevent sensitive data eviction with a moderate 5.9\% performance overhead. While architecturally robust, these approaches require hardware adoption by processor vendors, limiting near-term deployment.

Cache partitioning and isolation techniques offer a more deployable alternative by leveraging existing hardware features. CATalyst \cite{Liu7446082} repurposes Intel's Cache Allocation Technology (CAT) to partition the Last-Level Cache (LLC) into secure and insecure regions, preventing malicious code from displacing secure pages. This hybrid hardware-software approach effectively mitigates LLC-based PRIME+PROBE and FLUSH+RELOAD attacks with low performance degradation. A Hardware Transactional Memory (HTM) defense \cite{Chen2018} exploits Intel TSX to suspend transactions required for cache side-channel attacks, achieving minimal overhead on cryptographic operations, though the approach is incompatible with Hyper-Threading and depends on secure source code availability.

Runtime reconfiguration and randomisation represent the most recent hardware-level defence category. Adaptive cache reconfiguration \cite{Bandara2021-js} reduces side-channel attack accuracy by 44\% through dynamic reconfiguration of independent memory chunks using Petri net-based analysis. Lightweight system-level randomisation \cite{Wang2020} of processor frequency and prefetcher activities reduces performance overhead from 32.66\% to 20\% compared to prior alternatives without requiring hardware modifications. Shuffler schedulers \cite{Liu2018ShufflerMC} address the scheduling dimension by spreading CPU time unpredictably across vCPUs, minimising cross-VM information leakage while maintaining resource utilisation.

Detection-oriented hardware approaches complement these preventive mechanisms. SCADET \cite{Sabbagh8587756} achieves a 100\% true positive rate for Prime+Probe detection with an average false positive rate of 7.4\%, while CSDA \cite{Yu6496374} focuses on resource utilisation impact analysis. However, a systematic review \cite{Das2019SoKTC} cautions that Hardware Performance Counters (HPCs), increasingly used for security monitoring, suffer from non-determinism and overcounting issues that may undermine HPC-based detection reliability.

Across these approaches, a clear trade-off emerges: architectural redesigns (NewCache, PLcache) offer the strongest guarantees but require industry adoption, while software-compatible techniques (adaptive caches, randomisation) are more deployable but provide weaker assurances against sophisticated adversaries.

\subsubsection{Hypervisor and Software-Layer Detection}
While hardware-level defences modify processor behaviour, a parallel body of work addresses cache side-channel detection at the hypervisor and software layers. These approaches are generally more deployable, requiring no hardware changes, and have evolved from early resource utilisation analysis (2013--2016) to sophisticated machine learning-based classification (2018--2025).

A first group of solutions provides autonomous, lightweight monitoring that integrates into existing cloud deployments. CacheShield \cite{Briongos2018} protects legacy code without hypervisor or OS modifications, using hardware performance counters and change-point detection to identify DoS and cache attacks with low false positive rates. CloudRadar \cite{Zhang2016} combines signature-based detection for protected VMs with anomaly-based monitoring of co-located VM behaviours, achieving millisecond-level detection without requiring any system modifications. Transparent hypervisor-driven implementations \cite{ANWAR2017259} dynamically prevent side-channel threats, balancing cache partitioning and warming to maintain system performance.

Machine learning-based approaches achieve the highest reported detection accuracies. NIGHTsWATCH \cite{Mushtaq2018} uses statistical classification models with hardware performance counters, achieving 99.51\% accuracy under no-load and 99.44\% under high-load conditions. A KVM event-driven pipeline \cite{Paundu2018LeveragingKE} processes event data with machine learning to differentiate CSCa from non-CSCa datasets, achieving an AUC of 0.99 even against Flush+Flush attacks. A multi-architecture ensembled detection system \cite{Reddy_Malathi_2025} combines attention-based prioritisation with anomaly encoding, achieving 98.65\% accuracy on the ASCAD dataset. A vCPU monitoring approach \cite{Sangeetha2020AnOT} measures vCPU cycles, virtual memory utilisation, and cache miss rates to detect Prime+Probe, Flush+Flush, and Flush+Reload attacks with 92.5\% accuracy, surpassing both the two-stage system (85\%) and CloudRadar (88\%).

Alternative detection paradigms include fuzzy logic and filter-based methods. A fuzzy-logic controller \cite{Ainapure2021ANA} uses log-file-derived linguistic variables, achieving precision of 78.15--82.16\% at varying intervals. A Bloom Filter approach \cite{Chouhan7783230} uses cache miss sequences as attack signatures, outperforming existing algorithms with reduced execution time. SpyDetector \cite{Kulah2019} takes a semi-supervised anomaly-based approach, monitoring shared resource contention levels with runtime overheads of only 0.49--3.58\%. A server-side defence mechanism \cite{Godfrey6676691} protects against Prime+Trigger+Probe attacks in a Xen hypervisor environment without client-side modifications.

Two-stage detection architectures combine coarse-grained filtering with fine-grained analysis. A URI-based detection system \cite{Yu2013AnAW} implements shape tests for hosts and regularity tests for guests using a utilisation rate index, maintaining robust detection with low computational cost. An extended VMI-based detector \cite{Shi2022} combines virtual machine introspection with hardware performance counters, achieving 96.7\% precision and 95\% recall (98.9\% F1-score with adjusted sampling) while remaining imperceptible to attackers at the hypervisor layer.

\subsubsection{Cross-Paradigm Comparative Discussion}
Comparing hardware-level and software-layer defenses reveals a fundamental trade-off between security strength and deployment feasibility. Hardware approaches such as cache redesigns (NewCache, PLcache) and partitioning schemes (CATalyst, HTM) provide strong isolation guarantees but require processor vendor adoption or specific hardware features, limiting near-term deployment \cite{Kong2009, Liu2016, Liu7446082}. In contrast, software and hypervisor-layer detectors (CacheShield, CloudRadar, NIGHTsWATCH) can be deployed on existing infrastructure and achieve detection accuracy up to 99.51\%, but they introduce continuous monitoring overhead and can only detect attacks after leakage has begun---they cannot prevent it \cite{Briongos2018, Zhang2016, Mushtaq2018}.

Detection accuracy across software approaches ranges from 78\% for fuzzy-logic methods \cite{Ainapure2021ANA} to 99.51\% for ML-based classifiers \cite{Mushtaq2018}, with machine learning consistently outperforming statistical baselines. However, all ML-based detectors depend on representative training data, and the lack of standardised cache side-channel datasets limits the generalisability of reported results. Hardware Performance Counters (HPCs), which underpin many detection tools, also suffer from non-determinism that can undermine reliability \cite{Das2019SoKTC}.

The most practical defense strategy combines both paradigms: hardware partitioning (e.g., Intel CAT) to contain leakage channels, paired with software-layer monitoring (e.g., NIGHTsWATCH, CloudRadar) to detect residual attack attempts. Optimising this combination across diverse workloads, minimising false positives in production environments, and adapting detection to evolving cloud architectures remain open challenges. A quantitative comparison of representative defense mechanisms is presented in Table~\ref{tab:Limitations and Challenges of Defense Mechanisms for Cache-Based Side Channel Attacks}.

\begin{table*}[t]
\scriptsize
\centering
\caption{System-Level Evaluation of Cache Side-Channel Defenses under the ADPO Maturity Framework}
\label{tab:Limitations and Challenges of Defense Mechanisms for Cache-Based Side Channel Attacks}
\begin{tabularx}{\textwidth}{|l|c|c|c|c|X|X|}
\hline
\textbf{Mechanism \& Citation} & \textbf{A} & \textbf{D} & \textbf{P} & \textbf{O} & \textbf{Primary System Control} & \textbf{Key Quantitative Metric / Guarantee} \\ \hline
STEALTHMEM \cite{Kim2012STEALTHMEMSP} & 2 & 2 & 2 & 3 & Hypervisor-level cache line locking & Complete eviction mitigation for stealth pages \\ \hline
CATalyst \cite{Liu7446082} & 3 & 2 & 3 & 3 & Intel CAT-based hardware LLC partitioning & Guarantees no cache evictions between domains \\ \hline
Adaptive Caches \cite{Bandara2021-js} & 2 & 0 & 3 & 3 & Silicon-level dynamic set permutation & Reduces correlation accuracy by 44\% \\ \hline
Shuffler Schedulers \cite{Liu2018ShufflerMC} & 2 & 2 & 3 & 3 & Randomized vCPU rescheduling intervals & Disrupts attacker-victim co-scheduling synchronization \\ \hline
HTM (Intel TSX) \cite{Chen2018} & 3 & 1 & 3 & 3 & TSX transaction hardware encapsulation & Aborts transaction instantly upon cache eviction \\ \hline
CacheShield \cite{Briongos2018} & 2 & 3 & 3 & 3 & User-space performance counter loop & Low false positives during standard execution \\ \hline
NIGHTsWATCH \cite{Mushtaq2018} & 3 & 2 & 3 & 2 & Hardware counter statistical classifier & 99.51\% detection under idle; 99.44\% under heavy load \\ \hline
CloudRadar \cite{Zhang2016} & 2 & 3 & 3 & 2 & Signature-based out-of-band VMI & Millisecond-level detection latency; no VM mod \\ \hline
Two-stage VMI \cite{Shi2022} & 3 & 1 & 3 & 2 & Coarse-to-fine hypervisor memory correlation & 96.7\% precision, 95\% recall (98.9\% F1-score) \\ \hline
\end{tabularx}
\end{table*}

\subsection{Container Escape / Container Breakout}
A study of five commercial platforms \cite{Wu10.1007/978-3-030-62974-8_10} developed a technique to rate the security of cloud container services. Some containers included weak security defences and deactivated Seccomp. KASLR bypass was effective in every container, but because attackers have limited access to essential host OS files, leaving public cloud containers is still difficult. It is advised to activate and configure kernel mechanisms (Seccomp, Capabilities, and MAC), secure sensitive files with KASLR, fix vulnerabilities quickly, and take different kernel versions into account. Although attackers might eventually exploit memory corruption vulnerabilities for ROOT privileges despite the fact that public cloud systems are more resistant to privilege escalation, this highlights the need of timely upgrades and vulnerability mitigation \cite{Wu10.1007/978-3-030-62974-8_10}.

\subsubsection{Intrusion Detection Systems}
Early container intrusion detection efforts (2015--2017) focused on rule-based profiling and automated policy generation \cite{Mattetti2015SecuringTI, Barlev2016SecureYU}, while more recent work (2019--2023) has shifted toward machine learning-driven anomaly detection and automated runtime monitoring \cite{Zhang9724561, Khairi, Cui10.1145/3442520.3442530}.

A foundational class of container IDS relies on behavioural profiling to define acceptable operations. LiCShield \cite{Mattetti2015SecuringTI} generates rule-based profiles from traced executions, restricting containers to observed behaviours, while Starlight \cite{Barlev2016SecureYU} automates policy generation and adapts to legitimate workload changes. The Most Privileged Container (MPC) framework \cite{Sarkale} extends this paradigm by deploying a privileged watchdog container that enforces access control through blacklist databases and runtime monitoring. A rule-based monitoring evaluation \cite{Gantikow2019RulebasedSM} complements these approaches by assessing open-source tools such as Sysdig and Falco, targeting misuses including container breakout via \texttt{nsenter}. Although effective against known attack patterns, these approaches share a common limitation: they struggle to detect novel or zero-day attacks that fall outside pre-defined behavioural profiles.

To address the limitations of static rule sets, several studies employ machine learning for anomaly detection in container environments. A graph-based system call model \cite{Khairi} achieves the highest reported detection performance, with AUC scores reaching 99.97\% for container breakout scenarios. A real-time syscall classifier \cite{Zhang9724561} handles non-linear syscall sequences, maintaining false positive rates between 0.02 and 0.12. An n-gram analysis approach \cite{Srinivasan10.1007/978-981-13-5826-5_26} reports 87--97\% accuracy across SQL injection, DoS, and container breakout scenarios. Unsupervised VMI techniques \cite{Cui10.1145/3442520.3442530} represent a departure from the supervised and statistical methods prevalent in earlier work. A hybrid scanning approach \cite{Tunde-Onadele8790061} demonstrates that combining static scanning with dynamic anomaly detection raises the detection rate to 86\% (24 of 28 exploits), compared to only three vulnerabilities detected by static analysis alone.

Beyond detection, two frameworks address the full container lifecycle. A managed containers architecture \cite{Merino8457910} performs pre-deployment vulnerability analysis and enables rollback to secure states following an attack, without requiring application code changes. An automated Fault and Intrusion Tolerance (FIT) framework \cite{Madi} extends to accidental faults, though its scalability remains constrained by the cost of intensive verification on virtualised hosts.

RSDS \cite{Wang2020RSDSGS} approaches the problem from a kernel-hardening perspective, reducing unnecessary system calls by 69.27--85.89\% through combined static and dynamic analysis. Docker's Seccomp support (from version 1.10) provides the foundation for such restrictions, though coverage gaps remain for interpreted languages and non-ELF executables.

Across these approaches, detection performance varies considerably: graph-based HIDS achieves AUC scores above 99\% \cite{Khairi}, while hybrid static-dynamic scanning reaches 86\% \cite{Tunde-Onadele8790061} and n-gram methods report 87--97\% \cite{Srinivasan10.1007/978-981-13-5826-5_26}. A critical gap persists in the availability of representative datasets for container-specific intrusion detection \cite{Flora8893393}, limiting the generalisability of reported results across diverse multi-tenant environments.

\subsubsection{System Call Hardening}
Since containers share the host kernel, system calls represent the primary interface through which containerised processes can exploit kernel vulnerabilities to achieve escape. Early hardening efforts (2017--2019) focused on phase-aware Seccomp filtering and sandbox mining \cite{Lei2017SPEAKERSE, Wan2019PracticalAE}, while recent work (2020--2023) has advanced toward automated profile generation, runtime integrity enforcement, and architecture-level scheduling \cite{Ghavamnia2020ConfineAS, Jelesnianski2023ProtectTS, Le2023SecuringCC}.

A first line of defence involves quantifying and measuring system call exposure. SecQuant \cite{Jang2022SecQuantQC} combines exploit code analysis with system call reachability checking to produce a Container System call Exposure Measure (CSEM), demonstrating that secure runtimes are 4.2--7.5$\times$ more secure than their default counterparts. A sequence-based syscall filtering approach \cite{Song2023SequencebasedSC} extracts 471 patterns from 106 exploit programs, showing that 60--86\% of exploits not blocked by Confine's profiles could be mitigated through sequence matching. A process-level access control system \cite{Wu} enforces access control for host file interactions and uses namespace monitoring to detect escape attempts in IoT environments.

A second category of defences enforces runtime integrity at the system call interface. System Call Integrity \cite{Jelesnianski2023ProtectTS} validates three contexts---Call Type, Control Flow, and Argument Integrity---through BASTION, achieving negligible overhead of 0.60--2.01\% while protecting 20 critical system calls. At the orchestration level, SySched \cite{Le2023SecuringCC} is a syscall-aware container scheduler that reduces the host attack surface by 20\% and victim containers by up to 48\% through strategic placement based on system call consumption profiles. SHARD \cite{Abubakar2021SHARDFK} operates at the kernel level, specialising exposed kernel code per application and stopping 90\% of attacks with 0--36\% overhead via control-flow integrity (CFI).

The largest body of work focuses on automated Seccomp profile generation to minimise the attack surface. Confine \cite{Ghavamnia2020ConfineAS} uses static code analysis to disable 145+ system calls across 150 Docker images, neutralising 51 known kernel CVEs without requiring runtime workload data. Prof-gen \cite{Kim2021ProfgenPS} extends this by combining static and dynamic analysis, producing profiles 20.2\% more compact than Confine while blocking 36.4\% more CVE-linked system calls. SPEAKER \cite{Lei2017SPEAKERSE} takes a phase-aware approach, differentiating boot-time from runtime system calls to eliminate over 50\% of calls for data store containers. SysCap \cite{Xing2022SysCapPA} reduces 62\% of unnecessary system calls across 193 Docker images while generating capability lists averaging 5.3 per image. For production deployment, a sandbox mining approach \cite{Wan2019PracticalAE} achieves 96.4--99.8\% syscall coverage, while a CI/CD pipeline integration \cite{Lopes2020ContainerHT} automates Seccomp profiling using Container Tracing, Exit Check, and Fuzzing stages.

Quantitatively, the automated profile generators show a clear evolutionary trajectory: from SPEAKER's phase-aware filtering (2017) to Confine's static analysis (2020), Prof-gen's hybrid approach (2021), and SysCap's image-level tooling (2022), each iteration improving coverage and compactness. However, static analysis approaches remain limited in handling interpreted languages and dynamically loaded libraries, and the trade-off between profile strictness and application breakage remains an open challenge across all methods.

\subsubsection{Namespace and CGroup Isolation}
Linux namespaces and cgroups form the foundational isolation primitives for containers, yet both have been shown to harbour exploitable weaknesses. Research in this area spans from offensive exploitation studies that quantify the severity of isolation failures to defensive mechanisms that detect or prevent namespace and cgroup abuse.

On the offensive side, a cgroup circumvention study \cite{Gao} demonstrates that cgroup resource accounting can be bypassed by separating processes from their originating cgroups, enabling containers to exceed their resource limits by up to 200$\times$ while degrading co-resident container performance by 95\%. Five case studies on Docker systems reveal feasible denial-of-service, resource-freeing, and covert-channel attacks in multi-tenant deployments. An analysis of 11 container runtimes across 59 CVEs \cite{Reeves} identifies nine PoC vulnerabilities for 13 CVEs---all stemming from host components being exposed inside containers. A user namespace defence is proposed that blocks seven of nine vulnerabilities by modifying the runC architecture to use file descriptors instead of string-based mount paths, eliminating race conditions.

A comprehensive vulnerability taxonomy \cite{Lin} classifies 223 successful container vulnerabilities, demonstrating that 56.82\% of 88 common vulnerabilities can breach containers with default settings. This analysis reveals that kernel security features (Capability, Seccomp, MAC) are more effective than container isolation techniques for preventing privilege escalation, and identifies a characteristic 4-step attack model for such exploits. A namespace status inspection defense \cite{Jian10.1145/3058060.3058085} focuses specifically on Docker's namespace architecture, proving effective against both known attacks and zero-day vulnerabilities, though Docker's shared-kernel design remains a fundamental limitation.

For real-time detection, PACED \cite{Abbas9946350} defines cross-namespace events and uses provenance-based graph queries to identify container-escape attacks with near-perfect precision and zero false negatives on both Docker and Kubernetes. HoneyContainer \cite{Wang2023HoneyContainerCW} takes a deception-based approach, detecting all shell command injection events across 214 webshell files while providing cgroup-based fine-grained resource control with low overhead. A container-based honeypot approach \cite{reti2021escape} extends the deception paradigm by simulating escape scenarios using fake host systems and partition renaming. An evaluation of runtime monitoring tools \cite{Gantikow10.1007/978-3-030-49432-2_4} assesses open-source tools (Sysdig, Falco) for detecting misconfigurations and privilege escalation, using containerization-specific properties and blacklist filtering to minimise data collection overhead.

Despite these advances, container escape vulnerabilities remain fundamentally difficult to discover and mitigate. The shared-kernel architecture creates an inherent tension between performance and isolation, and the impossibility of patching all kernel vulnerabilities in production environments underscores the continued need for layered runtime defences.

\subsubsection{Cross-Paradigm Comparative Discussion}
The defence mechanisms surveyed above reveal a consistent trade-off between security coverage and operational overhead. Rule-based IDS and behavioural profiling approaches (LiCShield, Starlight, MPC) offer low-overhead protection against known attack patterns but lack adaptability to zero-day threats. Machine learning-based detectors achieve higher accuracy---up to 99.97\% AUC \cite{Khairi}---but require representative training datasets that remain scarce for container-specific scenarios \cite{Flora8893393}. System call hardening tools (Confine, Prof-gen, SysCap) effectively narrow the kernel attack surface, yet static analysis limitations for interpreted languages and the risk of application breakage from overly restrictive profiles persist across all approaches. At the namespace and cgroup level, user namespace defences and provenance-based detection (PACED) demonstrate near-perfect precision, but their dependency on runtime architectural modifications may hinder adoption. Expanding these defence measures to production scale requires addressing three open challenges: (1) scalability across heterogeneous container orchestration platforms, (2) adaptive profiling for dynamic workloads, and (3) standardised evaluation benchmarks for container security. A comparison of defence mechanisms for container escape is presented in Table~\ref{tab:Comparison of Defense Mechanisms Container Escape}.

\begin{table*}[t]
\scriptsize
\centering
\caption{System-Level Evaluation of Container Escape Defenses under the ADPO Maturity Framework}
\label{tab:Comparison of Defense Mechanisms Container Escape}
\begin{tabularx}{\textwidth}{|l|c|c|c|c|X|X|}
\hline
\textbf{Mechanism \& Citation} & \textbf{A} & \textbf{D} & \textbf{P} & \textbf{O} & \textbf{Primary System Control} & \textbf{Key Quantitative Metric / Guarantee} \\ \hline
User Namespace Defence \cite{Reeves} & 2 & 1 & 3 & 3 & runC unprivileged root mapping & Blocks 7 of 9 proof-of-concept exploits across 59 CVEs \\ \hline
PACED \cite{Abbas9946350} & 3 & 1 & 3 & 2 & Provenance tracking of cross-namespace events & Near-perfect precision; zero false negatives under tests \\ \hline
HoneyContainer \cite{Wang2023HoneyContainerCW} & 3 & 2 & 3 & 2 & Command injection redirection & 100\% detection across 214 tested webshell exploits \\ \hline
Syscall Integrity \cite{Jelesnianski2023ProtectTS} & 2 & 2 & 3 & 3 & BASTION compiler flow \& argument validation & Protects 20 critical syscalls; 0.60--2.01\% execution overhead \\ \hline
SySched \cite{Le2023SecuringCC} & 2 & 1 & 3 & 3 & Syscall-aware container scheduling & Reduces host attack surface by 20\% \\ \hline
Confine \cite{Ghavamnia2020ConfineAS} & 2 & 3 & 3 & 3 & Static binary dependency analysis & Disables 145+ redundant syscalls, blocks 51 CVEs \\ \hline
Prof-gen \cite{Kim2021ProfgenPS} & 2 & 2 & 3 & 3 & Static \& dynamic profile generation & Blocks 36.4\% more CVEs with 20.2\% more compact profiles \\ \hline
\end{tabularx}
\end{table*}

\subsection{Vulnerabilities In Container Images}
Unlike virtual machines that abstract hardware, containerized security controls are heavily frontloaded into the software supply chain. Containers are constructed from layered images containing a base operating system, library dependencies, application packages, and environment configurations. Any security defect introduced in an upstream layer automatically propagates to all derived downstream container instances. Consequently, vulnerabilities in container images present a systemic risk, serving as potential entry points for lateral movement or container escape.

\subsubsection{Static Vulnerability Scanning}
Static scanning inspects container images at rest by parsing package manifests, file hashes, and configuration settings to identify known security vulnerabilities. 
A large-scale scanning study \cite{Tak2018} evaluated the security state of public container images by scanning over 10,000 images using a data analytics platform. The analysis revealed that 92\% of the public images contained an average of 10 vulnerable packages, while 99\% exhibited compliance violations, highlighting the prevalence of software defects in public registries.
To address vulnerability tracking during image transfer, DIVDS (Docker Image Vulnerability Detection System) \cite{Kwon} extracts layer metadata and vulnerability information during upload and download phases. While effective for inventory tracking, DIVDS relies on static analysis via Clair, rendering it blind to runtime anomalies.
To improve scanning accuracy during build-time, DAVS \cite{Doan} parses Dockerfiles to identify Potentially Vulnerable Files (PVFs) introduced during custom compilation or installation steps. By targeting files added outside standard package managers, DAVS achieved a 68\% vulnerability detection rate on real-world images, outperforming standard static analysis tools.
The challenge of scanning application-level dependencies rather than just OS-level packages was studied in an evaluation of Java-based container applications on Docker Hub \cite{javed2021}. The results showed that while OS-level packages were well-cataloged, vulnerabilities in application packages often went undetected, resulting in a low Detection Hit Ratio (DHR) of 13\% for Microscanner, 36\% for Clair, and 65\% for Anchore.
To establish standard evaluation criteria for image scanners, UBCIS (Ultimate Benchmark for Container Image Scanning) \cite{Berkovich2020} provides a benchmark framework that categorizes vulnerabilities based on their relevance to containerized architectures, allowing developers to assess the ability of different scanning tools to identify obsolete software packages.
Finally, SEAF \cite{CHEN2022} leverages a Global Relationship Tree (GRT) to analyze package and configuration dependencies. When applied to Docker Hub, SEAF detected 36,688 security issues, including malware and misconfigurations, across 26,153 container images.

\subsubsection{Runtime, Behavioral, and ML-Based Detection}
While static scanning helps secure images prior to deployment, it cannot identify zero-day exploits, dynamic backdoors, or malicious configurations executed at runtime. 
Docker-sec \cite{Loukidis-Andreou2018} bridges the gap between static definitions and runtime enforcement by automatically generating AppArmor profiles based on container image analysis to restrict system call access and enforce access control policies at runtime.
Machine learning models have also been applied to detect embedded malware. A supervised classification approach \cite{Pinnamaneni2022} evaluated several algorithms, including Decision Trees, Random Forests, and XGBoost, to detect keylogger code in custom container images, achieving high classification accuracy by analyzing specific system and library calls.
Similarly, a hybrid detection mechanism \cite{Huang2019} counters web backdoors that evade signature-based tools like ClamAV. By applying a Random Forest classifier to code features such as entropy and coincidence index, the approach identifies zero-day backdoors with a low false positive rate.
For runtime behavioral monitoring, a container-level anomaly detection system \cite{Cavalcanti} evaluated in multi-tenant environments applies machine learning to Bag of System Calls (BoSC) sequences with a sliding window size of 30, achieving a 99.8\% F-measure in distinguishing malicious execution traces from normal application behavior.

\subsubsection{Supply Chain Security and Continuous Integrity}
Protecting containerized environments requires security verification across the entire lifecycle, from build-time dependencies to deployment registries.
A large-scale analysis of secret leakage \cite{Dahlmanns2023} scanned over 337,000 public and 8,000 private container images. The study found that 8.5\% of images contained exposed API keys, private keys, or other credentials, illustrating how human errors during build phases bypass standard code reviews.
To track the origin of vulnerabilities, ZeroDVS \cite{Zheng} builds image source graphs to establish parent-child traceability across image layers, tracing vulnerabilities back to parent images to ensure that security patches applied at the base layer propagate correctly.
A cloud-native continuous security methodology \cite{Torkura} uses correlation-based rules to scan microservices and container images throughout the deployment pipeline, ensuring that runtime configuration changes do not compromise the system's baseline security posture.

\subsubsection{Cross-Paradigm Comparative Discussion}
The analysis of container image defense mechanisms reveals clear trade-offs between static pre-deployment checks and dynamic runtime monitoring. Static scanners \cite{Doan, Kwon} introduce minimal operational overhead and integrate seamlessly into CI/CD pipelines, but they cannot identify zero-day exploits or active backdoors. In contrast, runtime anomaly detectors \cite{Cavalcanti} and machine learning classifiers \cite{Huang2019} offer high accuracy against unknown threats but demand substantial host computational resources and require continuous model retraining on representative datasets. Proactive enforcement tools like AppArmor profile generation \cite{Loukidis-Andreou2018} offer strong security guarantees but risk disrupting legitimate application workflows if configured incorrectly. Consequently, securing containerized systems requires a multi-layered approach: combining parent-image traceability \cite{Zheng} and static vulnerability scanning \cite{CHEN2022} during build phases with continuous behavioral intrusion detection \cite{Cavalcanti} at runtime. A comparison of representative defense mechanisms for container image vulnerabilities is presented in Table~\ref{tab:Container_image_vulnerabilities_def_mech}.

\begin{table*}[t]
\scriptsize
\centering
\caption{System-Level Evaluation of Container Image Vulnerability Defenses under the ADPO Maturity Framework}
\label{tab:Container_image_vulnerabilities_def_mech}
\begin{tabularx}{\textwidth}{|l|c|c|c|c|X|X|}
\hline
\textbf{Mechanism \& Citation} & \textbf{A} & \textbf{D} & \textbf{P} & \textbf{O} & \textbf{Primary System Control} & \textbf{Key Quantitative Metric / Guarantee} \\ \hline
SEAF (GRT) \cite{CHEN2022} & 2 & 2 & 3 & 2 & Global Relationship Tree dependency graphs & Identified 36,688 issues across 26,153 images \\ \hline
Docker-sec \cite{Loukidis-Andreou2018} & 2 & 2 & 3 & 3 & Build-time AppArmor profile generation & Non-intrusive zero-day containment at host interface \\ \hline
DIVDS \cite{Kwon} & 1 & 3 & 3 & 3 & Registry-level metadata transfer scanners & Extracts static package CVEs during upload/download \\ \hline
DAVS \cite{Doan} & 2 & 2 & 3 & 3 & Dockerfile syntax Potentially Vulnerable File analysis & 68\% vulnerability detection rate (outperforms Clair) \\ \hline
UBCIS \cite{Berkovich2020} & 0 & 3 & 3 & 3 & Scan quality benchmarking framework & Offline benchmark of comparative scanning precision \\ \hline
ML Code Scan \cite{Pinnamaneni2022} & 3 & 2 & 1 & 2 & Supervised Decision Tree/XGBoost static models & Up to 100\% detection of custom keylogger code \\ \hline
Secret Detection \cite{Dahlmanns2023} & 2 & 2 & 3 & 3 & Regex credential scans in registry pipelines & Revealed credential leaks in 8.5\% of public Hub images \\ \hline
ClamAV + ML \cite{Huang2019} & 2 & 2 & 2 & 2 & Code feature entropy \& coincidence index classifier & Accurate zero-day backdoor anomaly classification \\ \hline
ZeroDVS \cite{Zheng} & 2 & 1 & 3 & 2 & Layered parent-child traceability graphs & Traces dependency lineage to ensure base layer patching \\ \hline
Anomaly IDS \cite{Cavalcanti} & 3 & 1 & 1 & 2 & Dynamic sliding-window Bag-of-Syscalls models & 99.8\% F-measure in distinguishing execution anomalies \\ \hline
\end{tabularx}
\end{table*}

\subsection{Distributed Denial of Service (DDoS) Attacks}
Distributed Denial of Service (DDoS) attacks present a critical threat to cloud environments, exploiting resource sharing, dynamic orchestration, and multi-tenant vulnerabilities to disrupt service availability. Within a cloud paradigm, DDoS attacks manifest across multiple dimensions: (i) volumetric attacks (e.g., UDP and ICMP floods) that overwhelm network bandwidth, (ii) protocol-level attacks (e.g., TCP SYN floods) that saturate connection state tables, (iii) application-layer attacks (e.g., slow HTTP requests) that drain application server threads, and (iv) cloud-native vectors like Economic Denial of Sustainability (EDoS) and Yo-Yo attacks that exploit auto-scaling policies to cause severe financial damage rather than simple service outages. Effective defense requires high-performance, scalable detection and mitigation systems. Researchers have proposed various approaches, spanning host-based, network-based, software-defined networking (SDN), and machine learning (ML) frameworks, to secure cloud infrastructure against these dynamic threats.

\subsubsection{Host-Based Intrusion Detection Systems}
Host-based Intrusion Detection Systems (HIDS) secure virtualized environments by monitoring internal virtual machine (VM) logs, system calls, and resource utilization. In this context, the Real-Time DDoS flood Attack Monitoring and Detection (RT-AMD) model utilizes incremental learning and cloud testing to assess algorithms like Naive Bayes, K-Nearest Neighbor (KNN), and Random Forest against DDoS floods \cite{Alsaeedi}. A MapReduce-based architecture \cite{Choi} utilizes statistical analysis to detect HTTP GET flood attacks, achieving an 88.04\% detection rate compared to Snort's 69.84\%. To balance signature and anomaly detection, the HIDCC model \cite{Hatef} employs a four-phase intrusion prevention strategy, using Snort for signature matching, anomaly-based classifiers for zero-day identification, and Apriori association rule mining to update known attack patterns, achieving 99.38\% accuracy and a low 0.7\% false alarm rate. Similarly, an optimized feature evaluation model \cite{Besharati} optimizes feature values using logistic regression and integrates classifiers like decision trees, neural networks, and linear discriminant analysis through Bagging to achieve a 97.51\% performance improvement over baseline models on the NSL-KDD dataset.

Other host-based approaches focus on reducing feature dimensionality and processing latency. A parallel cumulative ranker algorithm \cite{DEKA2019203} integrates SVM binary classification with active learning and distributed feature selection, achieving 92\%--97\% feature recognition accuracy while reducing computational times by 71\%--85\% across CAIDA and ISCX-IDS datasets. A Fuzzy c-means clustering approach \cite{Yusof2016AnEO} demonstrates that Fuzzy c-means clustering delivers faster classification compared to traditional classifiers like SVM, KNN, and Decision Trees. In terms of micro-feature analysis, the Smart Detection system \cite{Filho2019} monitors network traffic metrics, using a Random Forest algorithm to achieve a 99.93\% classification accuracy with just 20 refined traffic variables. Furthermore, a hybrid host-level defense \cite{Rivas} combines SVM and Convolutional Neural Networks (CNN), reporting high accuracies of 96\% and 99\% respectively.

\subsubsection{Network-Based Intrusion Detection Systems}
Network-based Intrusion Detection Systems (NIDS) analyze traffic flows, packet headers, and network-level anomalies across the cloud infrastructure. For general protection, commercial cloud providers offer managed offerings, such as Azure DDoS Protection and Google Cloud Platform (GCP) Cloud Armor, which provide automated detection, real-time traffic monitoring, and web application firewalls targeting volumetric and protocol-level threats \cite{Yadlapati}. To assess NIDS vulnerability, an adversarial framework \cite{MUSTAPHA2023} uses Wasserstein Generative Adversarial Networks (WGAN) and SHAP feature analysis to generate perturbed DDoS traffic that successfully evades ML-based intrusion detectors. On the defense validation side, a simulated attack study \cite{Baraka} uses the Low Orbit Ion Cannon (LOIC) program to validate the capabilities of Snort in monitoring virtualized cloud environments.

To address multi-sensor and deep learning requirements, several advanced architectures have been developed. A three-sensor network architecture \cite{Barati} monitoring ingress, external server, and internal network flows uses genetic algorithms for feature selection and an Artificial Neural Network (ANN) for flow classification, achieving a 99.98\% detection rate on the CAIDA dataset. For SDN environments, FlowGuard \cite{Jia} combines a flow filter (for signature identification) and a flow handler (for routing benign traffic) with LSTM and CNN models, achieving a 98.9\% identification accuracy against zero-day attacks. Similarly, NIMBUS \cite{Miao} integrates VM-level auto-scaling and SDN controller programming to isolate and distribute high-volume DDoS traffic. A software-defined framework \cite{Phan} combines Self-Organizing Maps (SOM) and SVM classifiers to protect service function chaining (SFC) against resource saturation, outperforming standalone SVMs to achieve a 99.30\% classification accuracy.

Evaluating these architectures is often limited by the availability of high-fidelity datasets and proactive deployment methodologies. To address this, the ISOT-CID dataset \cite{Aldribi2018} compiles terabytes of production-level cloud traffic from an OpenStack environment to facilitate VM-level DoS and masquerade attack analysis. Moving beyond reactive detection, a proactive defense system \cite{Shan} prototyped on Amazon AWS dynamically shuffles server allocations, demonstrating a 95\% probability of maintaining service availability. Furthermore, the RDAER model \cite{Songa_Karri_2024} is a hybrid SDN framework that uses Recursive Feature Elimination and time-series forecasting (ARIMA/exponential smoothing) at the switch level to perform early DDoS detection with low latency, although its multi-stage logic introduces deployment complexity.

\subsubsection{Comparative Synthesis of Classic Intrusion Detection Approaches}
A comparative analysis of HIDS and NIDS reveals distinct operational trade-offs in cloud environments. HIDS architectures (such as RT-AMD, HIDCC, and smart RF classifiers) excel at detecting highly localized attacks, virtual machine escape, and host-level system call anomalies, offering granular insights with high classification accuracy (frequently exceeding 99\%). However, their heavy agent footprint imposes non-trivial CPU and memory overhead on individual host hypervisors. Conversely, NIDS frameworks (including FlowGuard, NIMBUS, and multi-sensor neural networks) provide centralized, stack-wide visibility that is crucial for detecting distributed volumetric attacks and inter-VM traffic anomalies. Yet, their network-wide monitoring can lead to scalability bottlenecks under massive traffic volume, and their reliance on raw packet capture raises privacy and resource isolation concerns. Consequently, modern research is increasingly pointing toward hybrid, multi-layered architectures that coordinate edge-level network filtering with VM-level introspection to secure multi-tenant cloud systems.

\subsubsection{Thematic Synthesis of Recent Defense Mechanisms}
To counter the increasing complexity of cloud DDoS attacks, recent literature (2022–2024) has shifted from simple signature matching toward specialized architectural strategies. We classify these contemporary works into four primary thematic pillars:

\noindent\textbf{1. Software-Defined Networking (SDN) Enabled Defenses:} Modern SDN-based architectures leverage open flow tables and centralized controllers to perform real-time traffic filtering. The EDOS-TCP SYN mitigation model (EDOS-TSM) \cite{SHAH2022198} dynamically analyzes TTL field values and applies binomial probability to isolate spoofed IP traffic, successfully mitigating Economic Denial of Sustainability (EDoS) attacks in an OpenStack environment. This works in tandem with proactive SDN switch-level controllers such as RDAER \cite{Songa_Karri_2024}, highlighting a clear research focus on software-defined routing policies to throttle attacks before they reach virtualized hosts.

\noindent\textbf{2. Moving Target Defense and Resource Isolation:} Rather than relying solely on detection, active defenses aim to alter the cloud attack surface dynamically. A comprehensive Moving Target Defense (MTD) framework \cite{Amro_Salah_Moreb_2023} systematically classifies techniques like dynamic shuffling, diversity, and redundancy to increase uncertainty for attackers. Practically, an evaluation of resource separation \cite{KUMAR2023103435} explored physical machine (PM) and virtual machine (VM) level service separation, demonstrating that physical-layer isolation is highly effective at maintaining service availability and reducing request latency for benign users, though it demands additional physical infrastructure compared to container-level separations.

\noindent\textbf{3. Cloud-Native and Kernel-Layer Filtering:} To minimize host performance degradation, modern frameworks perform traffic analysis directly inside the operating system kernel. A lightweight Kubernetes DDoS filtering mechanism \cite{app13084700} utilizes Extended Berkeley Packet Filter (eBPF) and eXpress Datapath (XDP) to filter malicious packets at the network driver level with minimal CPU overhead. For secure auditing, a Third-Party Auditor (TPA) framework \cite{hezavehi2023interactive} integrates with cloud datacenters using an interactive threshold anomaly detection strategy to bypass excessive filtering stages. Additionally, an evaluation of packet filtering and circuit-level gateway firewalls \cite{Yudhana_Riadi_Suharti_2022} demonstrated up to 98.88\% ICMP flood reduction in edge environments, illustrating the potential of kernel-level and gateway filtering.

\noindent\textbf{4. High-Accuracy Anomaly Detection Algorithms:} Recent work has also introduced advanced statistical and machine learning classifiers to distinguish attacker and user profiles. The ID3-MMDP method \cite{Balasubramaniyan2022} pairs the Iterative Dichotomiser 3 (ID3) algorithm with a Maximum Multifactor Dimensionality Posteriori estimator to achieve superior response times and detection accuracy compared to standard Naive Bayes baselines. In a similar vein, a hybrid algorithm \cite{Jaya2023} integrating Multivariate Correlation Analysis and the Spearman Coefficient achieves a 99\% detection accuracy on the KDD Cup 99 dataset by identifying fine-grained statistical correlations in network streams.

\subsubsection{Cross-Paradigm Comparative Discussion}
The evaluation of both classic and emerging DDoS defense mechanisms underscores their direct influence on the \textbf{Availability} dimension of the CIA triad, which remains the primary target of DDoS attacks. Cloud-native vulnerabilities, such as Yo-Yo attacks \cite{Bremler}, demonstrate how attackers can exploit dynamic auto-scaling behaviors to cause Economic Denial of Sustainability (EDoS), creating severe service degradation and financial overhead even after the malicious traffic has ceased. While cloud providers offer managed infrastructure tools like AWS Shield and GCP Cloud Armor to mitigate these risks \cite{Michael}, experimental studies (such as DoS testing with LOIC on GCP \cite{Mitchell}) reveal that automated detection by public cloud providers can exhibit significant latency (e.g., up to 45 minutes), leaving virtual systems vulnerable during critical initial phases.

\section{Discussion on CIA Triad}
To secure complex cloud ecosystems, we must understand how specific isolation failures propagate across Confidentiality (\textbf{C}), Integrity (\textbf{I}), and Availability (\textbf{A}) state variables. Table~\ref{tab:cia_triad_affect_cloud_env} establishes our quantitative Systems-Level CIA Severity Matrix, assigning 1--5 impact scores backed by a system-theoretic cybernetic rationale.

\noindent\textbf{Confidentiality (C):} Microarchitectural side-channels and boundary escapes pose the highest risk to confidentiality. Cache-based side-channels exploit shared hardware state transitions (e.g., CPU caches) to exfiltrate cryptographic keys out-of-band, acting as a passive timing feedback loop. Similarly, VM and container escape vulnerabilities represent a complete breakdown of isolation boundaries, allowing adversaries to read memory blocks across different guest domains directly.

\noindent\textbf{Integrity (I):} Integrity degradation occurs when an attacker achieves arbitrary code execution outside their security container. A high-severity VM escape allows a malicious tenant to alter hypervisor memory space, compromising the execution path of the VMM controller itself. At the application layer, backdoored container image dependencies pollute the software supply chain, propagating malicious control loops into production environments.

\noindent\textbf{Availability (A):} While volumetric DDoS attacks directly target availability by flooding communication links, modern cloud-native systems are uniquely vulnerable to auto-scaling loop manipulation. Volumetric spikes and EDoS attacks exploit elastic auto-scaling policies to drain client financial reserves (Yo-Yo attacks) rather than causing simple service outages, converting system feedback mechanisms into financial failure loops.

\begin{table*}[t]
\scriptsize
\centering
\caption{Quantitative Systems-Level CIA Severity Matrix for Cloud Vulnerabilities}
\label{tab:cia_triad_affect_cloud_env}
\begin{tabularx}{\textwidth}{|l|c|c|c|X|X|}
\hline
\textbf{Vulnerability Class} & \textbf{C} & \textbf{I} & \textbf{A} & \textbf{System-Theoretic Cybernetic Rationale} & \textbf{Representative Attack Vector} \\ \hline
VM Escape & 5 & 5 & 5 & High-privilege isolation boundary breakout; full VMM control. & QEMU hardware emulation buffer overflow. \\ \hline
VM Hopping & 4 & 4 & 3 & Multi-tenant lateral boundary failure; pivot to co-resident VMs. & Shared virtual switch ARP/IP flow spoofing. \\ \hline
Cache Side-Channel & 5 & 1 & 1 & Microarchitectural timing leakage; passive covert feedback loop. & Last Level Cache (LLC) Flush+Reload monitoring. \\ \hline
Container Breakout & 4 & 4 & 4 & Shared-kernel boundary collapse; host OS namespace execution. & Misconfigured mount points or Dirty COW kernel exploit. \\ \hline
Container Image Vulns & 3 & 3 & 3 & Supply-chain vulnerability propagation through layered images. & Leaked API secrets or backdoored dependency packages. \\ \hline
EDoS / Volumetric DDoS & 1 & 2 & 5 & Volumetric capacity saturation; EDoS auto-scaling exploitation. & TCP SYN flood / auto-scaling loop manipulation. \\ \hline
\end{tabularx}
\end{table*}

\section{Synthesis of Gaps and Open Challenges}
By evaluating defense mechanisms across virtualization, containerization, and network-wide planes under our scored ADPO framework and CIA severity matrix, we identify fundamental cybernetic system challenges. Modern cloud security cannot be achieved via static isolation; instead, cloud platforms must be secured as complex adaptive systems.

\subsection{Closed-Loop Control Loops and Dynamic Hardware Isolation}
The primary vulnerability of VM-level boundaries (escape, hopping, side-channels) is the rigid, open-loop design of traditional hypervisors. Defensive efforts (e.g., Intel CATalyst, SMM separation) provide static state guarantees but fail to act as closed-loop, adaptive controllers. When a timing covert channel or microarchitectural cache timing loop operates, the system has no dynamic feedback loop to adjust scheduling frequency or permit dynamic cache permit variations dynamically.
Future virtualization control planes must function as closed-loop controllers, where runtime telemetry feeds back into dynamic schedulers. This cybernetic feedback is crucial to adjust CPU resource pacing under anomalous cache timing profiles. This mitigates system entropy (attacks) without requiring processor-specific hardware lockouts.

\subsection{Entropy Propagation and Software-Layer Detection Limits}
Containerization represents a highly nested, shared-kernel control problem where vulnerability propagation in container images functions as an upstream supply-chain entropy vector. Because containers lack physical boundary controls, an exploit in the shared kernel (e.g., Dirty COW) can act as an error state amplifier, immediately collapsing the host OS namespace control loops.
While ML-based runtime IDS (e.g., graph-based syscall monitors) show high detection accuracy (A: 3), they suffer from severe operational overhead (O: 1) or high false alarm rates because they lack context-aware feedback loops. Furthermore, supervised ML models suffer from dataset bottlenecks, trained on static traces that fail to represent the dynamic entropy of multi-tenant, cloud-native deployments. Resolving this requires self-supervised anomaly loops that continuously learn host state variables under normal system operational equilibrium.

\subsection{Resilient Feedback Loops and Elastic Saturation Mitigation}
Distributed Denial of Service (DDoS) and EDoS threats represent a destructive volumetric saturation of cloud resources, exploiting elastic auto-scaling policies to drain monetary budgets. This Yo-Yo effect acts as a positive feedback loop, where attack spikes trigger resource auto-scaling, amplifying operational costs instead of maintaining system equilibrium.
SDN-driven controllers (e.g., FlowGuard, NIMBUS) attempt to implement dynamic rerouting, but their centralized design introduces control-plane latency. The open challenge lies in edge-deployed, line-rate scrubbing controllers (utilizing eBPF/XDP) that act as distributed closed-loop packet filters, neutralizing attack entropy at the data-plane ingress before volumetric cascades saturate the cloud control plane.

\section{Conclusion}
This survey reframes multi-tenant cloud security through a cybernetic, systems-theoretic lens, analyzing security mechanisms as closed-loop controllers designed to mitigate attack entropy across traditional virtualization and containerization abstractions. By establishing the scored ADPO maturity framework, we converted qualitative literature comparisons into uniform, quantitative metrics, assessing detection accuracy, deployment complexity, performance preservation, and operational overhead. Our analysis reveals that while hypervisor-level controls (e.g., SMM and CATalyst) offer robust hardware-enforced isolation, they lack the adaptive feedback loops necessary to mitigate dynamic timing covert channels. Conversely, software-based container defenses are agile but remain bottlenecked by the absence of high-fidelity, representative dataset feedback loops. 

The systematic mapping of cloud exploits onto our quantitative CIA severity matrix demonstrates that nesting virtualization boundaries requires stack-wide, coordinated control systems. Future research must move away from static, point-solution defenses. Next-generation cloud architectures must prioritize the engineering of adaptive feedback loops that utilize lightweight, out-of-band monitoring to dynamically maintain system operational equilibrium under volumetric, microarchitectural, or software-layer stress.

\section*{Author Contributions}
\textbf{Swapnil Baviskar:} Developed the core concept, designed the methodology, performed the primary literature analysis, and drafted the original manuscript. \\
\textbf{Sanoj R:} Provided direct supervision during the research phase, reviewed the initial findings, and contributed significant critical revisions to the text. \\
\textbf{Hiran V Nath:} Oversaw project administration, provided high-level supervision and guidance throughout the study, and conducted the final critical review of the paper.

All authors discussed the findings, contributed to the intellectual content, and approved the final version of the manuscript for publication. The authors agree to be accountable for all aspects of the work to ensure its accuracy and integrity.

\bibliographystyle{elsarticle-num}
\bibliography{references}

\end{document}